\documentclass[%
 reprint,
superscriptaddress,
 amsmath,amssymb,
 aps,showkeys,
]{revtex4-1}

\usepackage{xspace}
\usepackage{amsmath}
\usepackage{color}
\usepackage{graphicx}% Include figure files
\usepackage{dcolumn}% Align table columns on decimal point
\usepackage{bm}% bold math
\usepackage{hyperref}
\usepackage{float}

\begin{document}

%title
\title{Interaction-induced zero-energy pinning and quantum dot formation in Majorana nanowires}

\author{Samuel D. Escribano}
\affiliation{Departamento de F{\'i}sica de la Materia Condensada C3 and}
\author{Alfredo Levy Yeyati}
\affiliation{Departamento de F{\'i}sica Te{\'o}rica de la Materia Condensada C5, 
Condensed Matter Physics Center (IFIMAC) and Instituto Nicol\'as  Cabrera,
 Universidad Aut{\'o}noma de Madrid, E-28049 Madrid, Spain}
 \author{Elsa Prada}
 \email[Corresponding author: ]{elsa.prada@uam.es}
 \affiliation{Departamento de F{\'i}sica de la Materia Condensada C3 and}

\begin{abstract}
Majorana modes emerge in non-trivial topological phases at the edges of some specific materials, like proximitized semiconducting nanowires under an external magnetic field. Ideally, they are non-local states that are charge neutral superpositions of electrons and holes. However, in nanowires of realistic length their wave functions overlap and acquire a finite charge that makes them susceptible to interactions, specifically with the image charges that arise in the electrostatic environment. Considering a realistic three-dimensional model of the dielectric surroundings, here we show that, under certain circumstances, this interaction leads to a suppression of the Majorana oscillations predicted by simpler theoretical models, and to the formation of low-energy quantum dot states that interact with the Majorana modes. Both features are observed in
recent experiments on the detection of Majoranas and could thus help to properly characterize them. 
\end{abstract}


 \maketitle

%%%%%%%%%%%%%%%%%%%%%%%%%%%%%%%%%%%%%%%%%%%%%%%%%%%%%%%%%%%%%%%%%%%%%
%%main text
\section{Introduction}
Semiconducting nanowires with strong spin-orbit interaction, like InAs or InSb, are becoming ideal systems for the artificial generation of topological 
superconductivity \cite{Alicea:RPP12,Beenakker:ARCMP13,Aguado:RNC17}. In addition to its fundamental interest, such nanowires, that may host Majorana bound states (MBSs) at their ends or interfaces \cite{Oreg:PRL10,Lutchyn:PRL10}, constitute promising platforms for Majorana-based
quantum computing devices \cite{Kitaev:AOP03,Nayak:RMP08,Aasen:PRX16,Lutchyn:17}.
Progress in fabrication techniques have allowed to induce a {\it hard} 
superconducting gap in InAs \cite{Albrecht:N16} or InSb \cite{Gazibegovic:N17} nanowires with epitaxially deposited 
Al layer. Moreover, last generation devices exhibit
a very low degree of disorder which allows them to almost reach the
ballistic limit \cite{Zhang:NC17,Nichele:PRL17,Zhang:17}. 

In spite of these advances, the experimental signatures of MBSs in
the nanowire devices deviate significantly in several aspects from the
theoretical predictions of minimal models. This is the case, for instance, regarding 
the behavior of the subgap 
conductance through the proximitized nanowire, which
has been addressed in several experiments \cite{Mourik:S12,Das:NP12,Deng:NL12,Churchill:PRB13,Deng:S16,Albrecht:N16,Nichele:PRL17,Zhang:NC17,Zhang:17}. 
In a long wire (whose length is much greater than the induced coherence length) the presence of MBSs manifests in the 
appearance of a zero bias
conductance peak whose width is controlled by the normal state 
conductance \cite{Zazunov:PRB16}. However, for the
typical wire lengths explored in actual experiments (which are of the
order of a few $\mu m$) it is expected that the overlap between MBSs
located at both ends of the wire gives rise to conventional Andreev bound
states which deviate from zero energy, leading to an oscillatory pattern
in the low bias conductance as a function of Zeeman field, chemical potential or wire length \cite{Prada:PRB12,Das-Sarma:PRB12,Rainis:PRB13}.  
Conspicuously, in most

\begin{figure}[H]
\begin{center}
\includegraphics{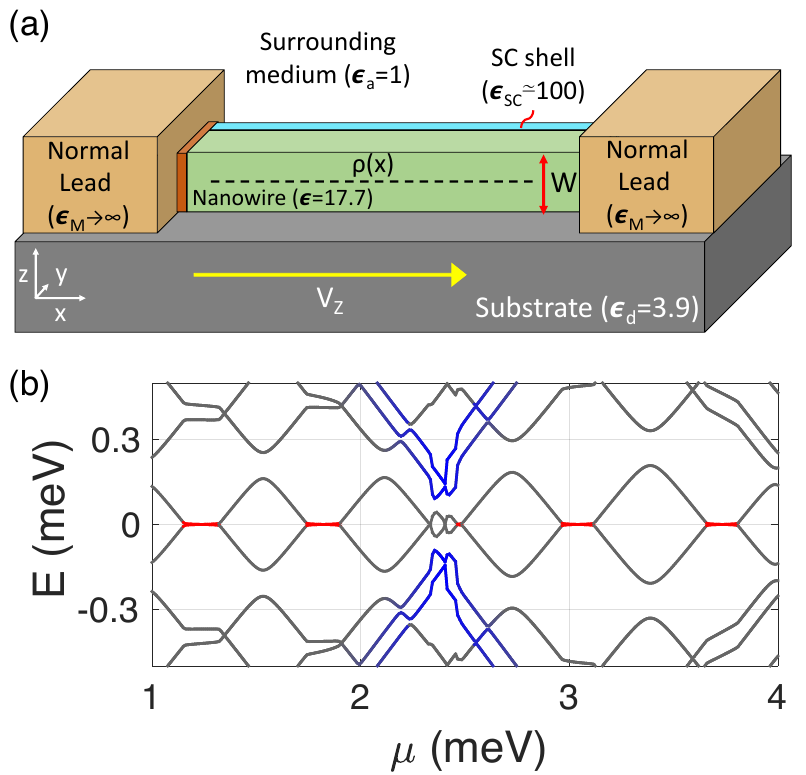}
\end{center}
\caption{(a) Schematic representation of the setup analyzed in the present work.
A nanowire of rectangular cross-section (green) lying on an insulating substrate (grey) and in contact to a thin metallic layer in one of its faces (light blue), corresponding to the parent superconductor, and
two normal-metal leads at its ends (orange) separated by tunnel barriers (brown). Typical values for the dielectric constants for
each region are indicated. (b) Low energy spectrum versus chemical potential $\mu$ for a wire of thickness $W=100$nm and length $L=1\mu$m. Other parameters are the spin-orbit coupling $\alpha=20$nm$\cdot$meV, the induced pairing energy $\Delta=0.3$meV and the Zeeman energy $V_Z=2$meV. Electrostatic environment-induced zero energy pinned regions between Majorana oscillations are indicated in red. Quantum dot levels (in blue), originated at the wire's edges due to the interaction with the bulk contacts, anticross with Majorana levels and remove their zero-energy pinning.}
\label{fig1}
\end{figure}

\noindent of the available
experimental data the emergence of a robust zero bias conductance peak is observed above
some critical Zeeman field without the expected
oscillatory pattern \cite{Deng:S16,Zhang:NC17,Vaitiekenas:17,Deng:17}. Several mechanisms have been proposed to account for the reduction or lack of oscillations, such as smooth confinement \cite{Kells:PRB12,Prada:PRB12,Liu:PRB17a,Moore:17}, strong spin-orbit coupling \cite{Liu:PRB17}, position dependent pairing \cite{Fleckenstein:17}, orbital magnetic effects \cite{Dmytruk:17}, Coulomb repulsion
among the carriers in the nanowire \cite{Das-Sarma:PRB12}, or the presence of the normal drain lead connected to the hybrid wire \cite{Danon:PRB17}.

Another source of Majorana oscillation suppression was put forward by some of us in a recent work \cite{Dominguez:npj17}. The key realization is that MBSs in a finite-length wire posses a finite charge, typically distributed uniformly along the wire \cite{Ben-Shach:PRB15}, which can be susceptible to electrostatic interactions with the surrounding medium. We considered the case of a grounded
parent superconductor, thus avoiding the effect of a charging energy
associated to the Cooper pairs, and showed that, in such case, a residual
effect of interactions may arise from the image charges induced in the 
electrostatic environment of the nanowire. Using a simple model for
the induced potential we concluded that, in typical experimental setups,
interactions would lead to a {\it pinning} of the MBSs to zero energy around parity crossings and,
thus, to more robust zero bias conductance peaks than predicted
by the non-interacting models.

The aim of the present work is to test the validity of the predictions of Ref. \cite{Dominguez:npj17} 
for the case of more realistic calculations of the induced 
electrostatic potential, taking into account the actual three dimensional
(3D) geometry as well as the effect of nearby metallic
leads. We consider the geometry depicted in Fig. \ref{fig1}(a), where
a nanowire of rectangular cross section lies on an insulating substrate
(typically ${\mathrm{SiO_{2}}}$) and is
contacted to a thin superconducting (SC) layer on one of its faces and to 
two bulk normal leads at both ends, separated by thin insulating barriers. In Fig. \ref{fig1}(a) we indicate
the characteristic dielectric constants of each region, which are relevant
for the calculation of the induced potential through the Poisson's equation (discussed bellow). 
Our aim is to solve this equation together with the Bogoliubov de Gennes
equation for determining self-consistently
the charge density $\rho(x)$ along the nanowire. For this purpose we derive
a generalized method of image charges which allows us to calculate the induced 
potential in rather general conditions, taking into account a 3D electrostatic
environment as the one shown in Fig. \ref{fig1}(a).

We find two main effects coming from this interaction and exemplified in Fig.\ref{fig1}(b). One is, as stated before, the suppression of Majorana oscillations around parity crossings (zero energy crossings where the total fermion parity of the wire changes), both as a function of the Zeeman energy $V_Z$ and the wire's chemical potential $\mu$. This effect is produced because, at each parity crossing, a finite Majorana charge $Q_M$ enters the wire from the reservoir in an abrupt fashion. If the electrostatic screening is smaller inside the wire than in the reservoirs, a repulsive interaction is produced between the incoming charge and its images, preventing its entrance. This translates into finite regions in parameter space [in red in Fig.\ref{fig1}(b)] where Majorana modes are \emph{pinned} to zero energy within a finite range of $V_Z$ or $\mu$ proportional to the Majorana charge $Q_M$ and the strength of the interaction. This was already shown in Ref. \cite{Dominguez:npj17} but for a simplified dielectric profile where the presence of the superconducting shell had been ignored. We here include it and find that the size of the pinned regions decreases but the pinning effect is still present under certain conditions that we discuss in detail below. Moreover, we explain the incompressible behavior of the electron liquid within these pinned regions in terms of the Majorana wave functions and their charge.
% (i.e. the dielectric constant of the wire is larger than that of its surroundings)

Another important effect of the electrostatic environment unexplored before is the creation of deep potential wells at the ends of the wire close to the bulk metallic electrodes. These wells, obtained explicitly here through the self-consistent calculation, are similar to the confinement potentials typical of quantum dots. Localized quantum dot-like energy levels in these regions disperse with magnetic field (or chemical potential) and appear below the induced gap in the wire spectrum [in blue in Fig.\ref{fig1}(b)]. In the topological regime, dot-like levels interact with Majorana states, anticrossing them when they approach zero energy. Similar phenomenology can be observed in some experiments \cite{Deng:S16,Zhang:17}, and has been likely found in other occasions but discarded by experimentalists looking for the simpler picture. Interestingly, it has been shown that the shape of these anticrossings can be used to quantify the degree of Majorana wave function non-locality \cite{Prada:PRB17,Clarke:PRB17}, a prediction that has been experimentally demonstrated recently \cite{Deng:17}. Here, we show that if the dot-Majorana levels interaction occurs in a pinning region, Majorana levels are forced to depart from zero energy, revealing the existence of a finite wave function overlap between them in spite of their zero energy. We analyze this behavior again in terms of the Majorana (and dot) wave functions and their charge.

The paper is organized as follows: in the following section (Sec. \ref{Section:Model}) we provide insight into the theoretical model used to treat interactions. In the next section (Sec. \ref{Section:no-leads}) we analyze the case in which the influence of the bulk normal leads can be neglected, recovering the pinning effect found in Ref. \cite{Dominguez:npj17} for a {\it repulsive} electrostatic
environment. However, we focus here on the electrostatic environment effects on the Majorana wave function, rather than on its spectral properties. In the next section (Sec. \ref{Section:with-leads}) we study the effect of including the bulk normal leads of Fig.\ref{fig1}(a), finding that they give rise to the formation of quantum dot-like bound states. We further analyze the interplay of such states with the MBSs. Finally, we present the conclusions of our work at the end (Sec. \ref{Section:Conclusions}). The robustness of the pinning effect is analyzed in detail in Supporting Information (SI) \hyperref[Section:SI4]{4}.

\section{Model and theoretical approach}
\label{Section:Model}

We model the electronic states along the proximitized Rashba
nanowire of length $L$ using the following single channel Hamiltonian \cite{Lutchyn:PRL10,Oreg:PRL10}

\begin{eqnarray}
H = \frac{1}{2} \int dx \Psi^{\dagger}(x) {\cal H} (x) \Psi(x), \quad \quad \quad \quad \quad\nonumber\\
{\cal H} = \left[-\frac{\hbar^2}{2m^*} \frac{\partial^2}{\partial x^2} - \mu 
+ e\phi(x) \right]\sigma_{0}\tau_z - i\alpha\sigma_y\tau_z \frac{\partial}{\partial x} +\nonumber\\
+V_Z\sigma_x\tau_z + \Delta\sigma_y \tau_y ,\quad
\label{model-Hamiltonian}
\end{eqnarray}
where $\Psi^{\dagger} = \left(\psi^{\dagger}_{\uparrow},\psi^{\dagger}_{\downarrow},\psi_{\uparrow},\psi_{\downarrow}\right)$ is a Nambu bi-spinor, $\psi_{\uparrow,\downarrow}(x)$ are electron annihilation operators, and $\sigma$ and $\tau$ are the Pauli matrices in spin and Nambu space, respectively. The model is defined by setting the parameters $m^*$, $\mu$,
$\alpha$, $V_Z$ and $\Delta$, corresponding to the effective mass, the chemical potential,
the spin-orbit coupling, the Zeeman energy caused by an external magnetic field, and the induced SC pairing potential. 

In Eq. (\ref{model-Hamiltonian}) we also include the electrostatic potential 
$\phi(x)$ felt by charges in the nanowire, which can be decomposed as
$\phi(x) = \phi_{int}(x) + \phi_b(x)$, where $\phi_{int}$ is the potential that arises from the free charges inside the nanowire, while $\phi_b$ corresponds to the potential created by
bound charges that emerge in the electrostatic environment.
%  As we argued in
%Ref. \cite{Dominguez:npj17}, the intrinsic contribution in the Hartree-Fock approximation has a negligible effect on the low-energy spectrum of the system, and in particular has no effect on the zero-energy crossings and their pinning. 
%For this reason we will concentrate solely on the effect of $\phi_b(x)$. 
We compute the electrostatic potential using the Poisson equation
\begin{eqnarray}
\vec{\nabla} \cdot \left[\epsilon(\vec{r})\vec{\nabla}\phi(\vec{r})\right] &=& \left< \rho(\vec{r})\right>,
\label{model-Poisson}
\end{eqnarray}
where $\epsilon(\vec{r})$ is the non-homogeneous dielectrical permittivity of the entire system  and $\left< \rho(\vec{r})\right>$ is the quantum and thermal average of the charge density of the nanowire obtained with Eq. (\ref{model-Hamiltonian}). The intrinsic part $\phi_{int}(x)$ of the potential satisfies an analogous equation with a uniform $\epsilon$ equal to that of the nanowire. The geometry depicted in Fig. \ref{fig1}(a) is taken into account through a piecewise $\epsilon\left(\vec r\right)$ function where each material is characterized by a different dielectric constant, so that $\epsilon\left(\vec r\right)$ changes abruptly at the interfaces. Then, assuming that the charge density in the nanowire is located along its symmetry
axis (x-axis), we obtain the electrostatic potential $\phi_{b}(x)$ using the method of image charges, as explained in detail in SI  \hyperref[Section:SI1]{1}. More precisely, $\phi_b$ is given by
$\phi_b(x) = \int dx V_b(x,x') \left< \hat{\rho}(x')\right>$, where $V_b(x,x')$ is a kernel determined in order to satisfy the proper boundary conditions. We find analytical expressions for
$V_b(x,x')$. They are simple but rather lengthy and are given in the SI
for two different cases: neglecting the effect of the bulk normal leads at the wire ends and including it. The results for these two cases are analyzed in the following sections.

The obtained potential $\phi(x)$ on the nanowire axis should be plugged back into Eq. (\ref{model-Hamiltonian}). The combined Poisson-Schr\"odinger problem must then be iterated until it achieves self-consistency. As shown in Ref. \cite{Dominguez:npj17}, the $\phi_{int}(x)$ part of the electrostatic solution (i.e. the intrinsic electron-electron interaction part of the problem), treated at the Hartree-Fock level, has a negligible effect on the low energy spectrum in the topological regime. We may therefore concentrate only on the self-consistency with $\phi_{b}(x)$. In SI \hyperref[Section:SI2]{2} we explain in detail the  self-consistent numerical method used to compute the electrostatic potential profile as well as the eigenvalues and eigenvectors of Eq. (\ref{model-Hamiltonian}). For completeness, in SI  \hyperref[Section:SI3]{3} we also show the effect of including the intrinsic interaction from $\phi_{int}(x)$, proving that its effect is small and that the main contribution stems from $\phi_{b}$. 

In the following calculations, we consider the dielectric constants shown in Fig. \ref{fig1}(a): for the dielectrics materials (the wire, the substrate and the surrounding medium), we use typical values \cite{Perry:11} ($\epsilon=17.7$, $\epsilon_{d}=3.9$ and $\epsilon_{a}\simeq1$, respectively); while for the metallic leads we assume that, because they are bulky, they screen  external electric fields perfectly (i.e. $\epsilon_{M}\rightarrow\infty$). This may not be the case for the SC shell, whose capability for screening external electric fields may be weaker due to its small thickness and unavoidable presence of possible disorder \cite{Hu:OL16}. If this is the case, it is then characterized by a \emph{finite} effective dielectric permittivity which depends on the SC shell width as well as its composition, as we show in SI \hyperref[Section:SI1]{1}. Some experiments \cite{Hovel:PRB10} have reported that for ultrathin metallic layers ($\sim5-10$nm) it is of the order of $\epsilon_{SC}\simeq100$. For these values, as we show in the next section, we find a repulsive environment, i.e., an environment whose effective permittivity is smaller than that of the wire so that the bound charges that arise at the interfaces have in average the same sign as the free charges.
%{\color{red}Because realistic dielectric constant values for epitaxially growth thin layers are not actually known, we have studied} 
We consider in SI \hyperref[Section:SI4]{4} the generality of our results versus the SC dielectric constant and the location of the charge density within the nanowire section. In Fig. 4(c) we show that, when the charge density is fixed at the center of the wire, as ε SC becomes larger the dielectric environment turns into an attractive one and the pinning effect is eventually lost. This however strongly depends on the location of the charge density. If, as pointed out in Ref. \cite{Knapp:17}, it happens to be close to the SC shell, the screening effect is larger and the pinning is suppressed. Nevertheless, as we analyze in Fig. 4(e), even if ε SC → ∞, the pinning effect survives when the wave function is located further away from the SC.

\section{Results and discussion}
\subsection{Results without bulk normal leads}
\label{Section:no-leads}

\begin{figure}
\begin{center}
\includegraphics{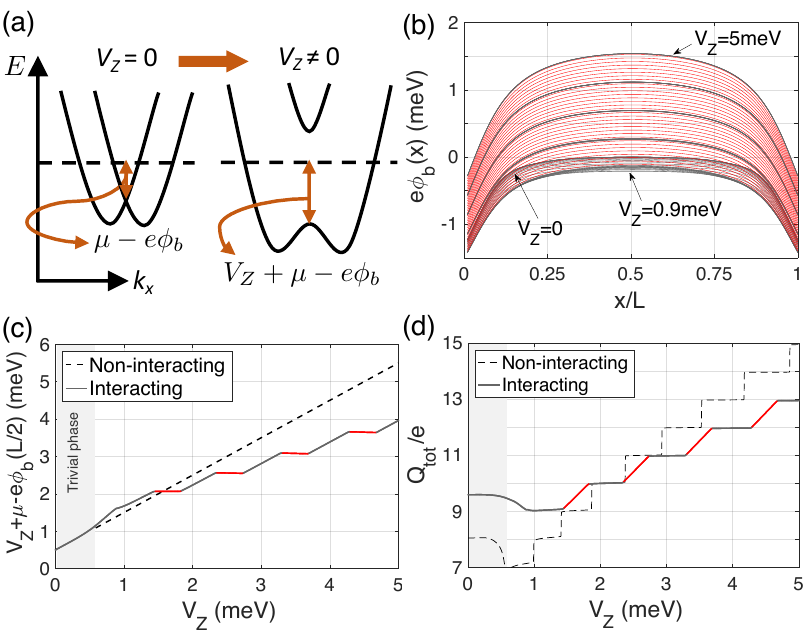}
\end{center}
\caption{Majorana nanowire subject to interactions from the electrostatic environment (ignoring the influence of the bulk normal leads at its ends). (a) Schematic of the nanowire's dispersion relation in the absence and in the presence of the Zeeman field. (b) Self-consistent induced potential energy $e\phi_b(x)$ along the wire's length for increasing values of the Zeeman splitting. Wire parameters as in Fig. \ref{fig1}(b) and with $\mu=0.5$meV. (c) Energy difference between the Fermi level and the band bottom at the center of the nanowire, $V_{Z}+\mu-e\phi_b(L/2)$, and (d) total charge $Q_{tot}$ of the nanowire as a function of $V_Z$ for the non-interacting (dashed) and interacting (solid line) cases. Red curves highlight parameter regions for which there is interaction-induced zero-energy pinning in the spectrum.}
\label{fig2}
\end{figure}

It is convenient to start by analyzing the simpler case in which we neglect the 
effect of the bulk normal leads in the induced potential $\phi_b$. As an example
we consider a nanowire of width $W=100$nm, length $L=1\mu$m and the following choice of realistic parameters: $m^*=0.015m_{e}$, $\alpha=20$nm$\cdot$meV, $\Delta=0.3$meV, $\mu=0.5$meV and $T=10$mK. These could correspond, for example, to a InSb nanowire in contact to an Al superconducting shell \cite{Zhang:17}, but similar results are obtained for InAs wire parameters \cite{Deng:S16}. For an infinite wire, a schematic representation of the energy bands is shown in Fig. \ref{fig2}(a) in the absence and in the presence of the Zeeman field. At zero temperature, the occupied states below the Fermi level are those between the horizontal dashed line and the band bottom. Apart from a small contribution coming from the spin-orbit energy, the position of the band bottom is controlled by the wire's chemical potential $\mu$, the Zeeman energy $V_Z$ and the induced potential energy $e\phi_b$. The magnetic field lowers the band bottom, charging the wire, whereas the induced potential energy, coming from electrostatic repulsion, tends to compensate that trend. In the finite-length wire, the evolution of the induced potential profile along the nanowire length ($x$-axis) for different Zeeman fields is shown in Fig. \ref{fig2}(b). As can be observed, the induced potential tends to expel charge from the center of the wire, where it is positive, while it bends downwards at its ends. On the other hand, the evolution of the potential with 
Zeeman field exhibits a step-like behavior with regions where it increases linearly with 
$V_Z$ (red curves), screening the magnetic field effects, and regions where it remains almost constant as $V_Z$ increases (grey curves). This causes the electron fluid to behave in an incompressible or compressible manner, respectively. This different behavior can be clearly seen in Fig. \ref{fig2}(c) where the electrochemical potential at the center of the wire, given by $V_Z+\mu-e\phi_{b}(L/2)$, is plotted against Zeeman splitting, both in the presence and absence of interactions. 
\begin{figure*}
\includegraphics{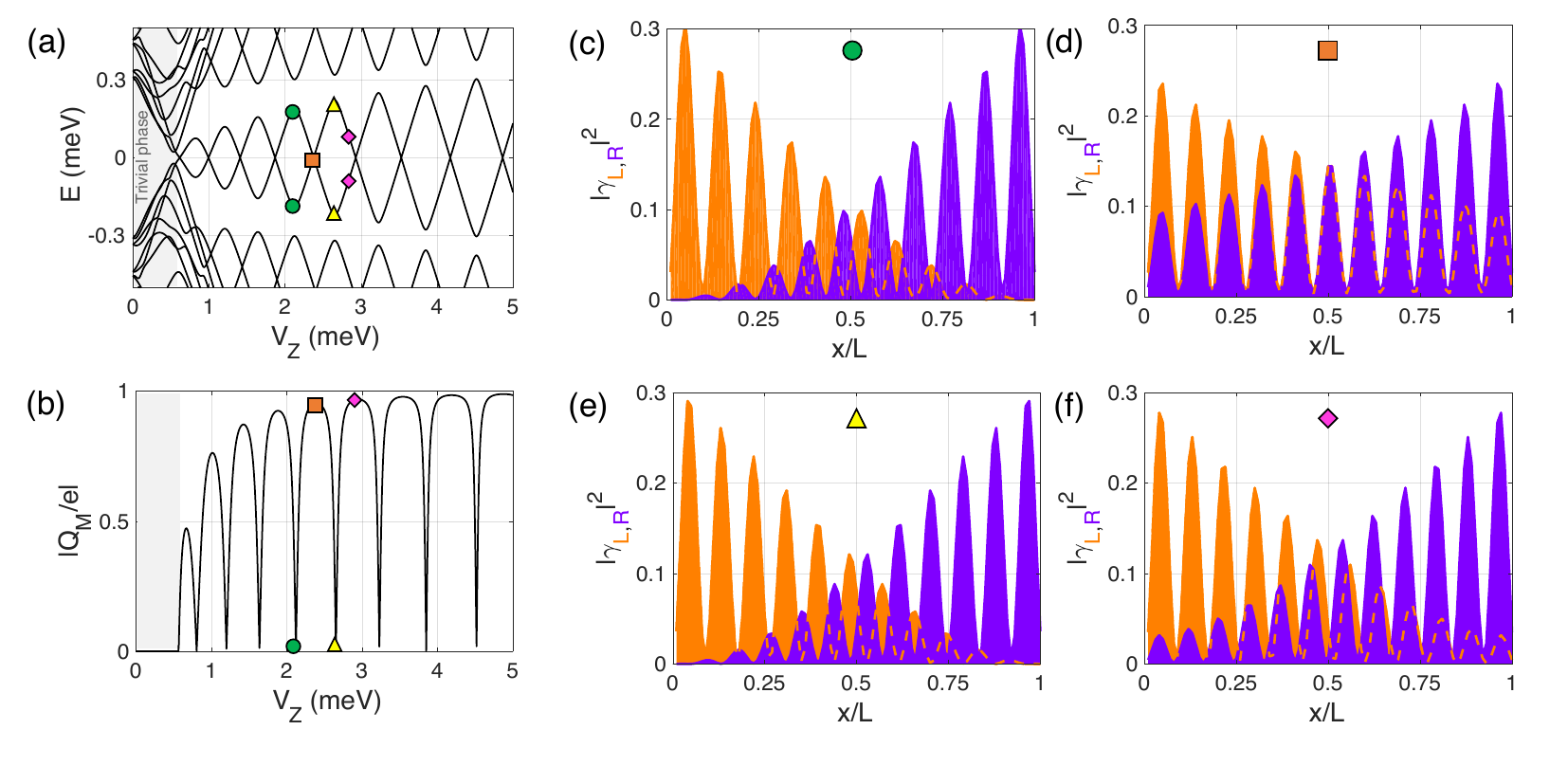}
\caption{Majorana wave functions in the non-interacting case: Energy levels (a) and the absolute value of the Majorana charge $Q_M$ (b) vs Zeeman energy. Panels (c-f) show the wave-function probability profiles of the two lowest energy states in the Majorana basis at selected values of the Zeeman field within the topological region. When the splitting is maximum (green circles and yellow triangles) the left and right Majorana wave-function oscillations are out of phase, whereas when the splitting is zero (orange square) they are in phase.}
\label{fig3}
\end{figure*}

\begin{figure*}
\includegraphics{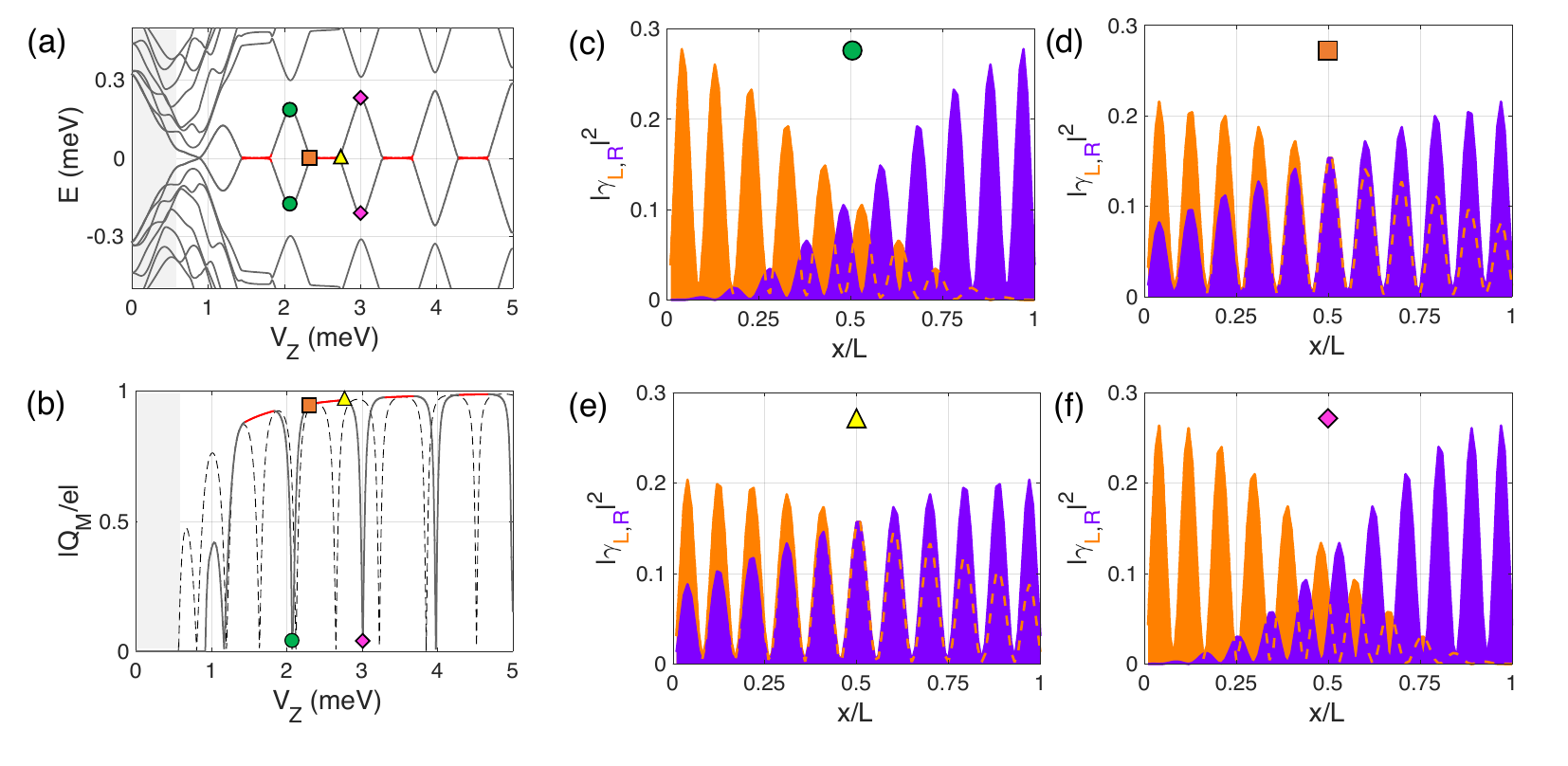}
\caption{Same as Fig. \ref{fig3} but for the interacting case (without leads). In the pinned regions the Majorana wave functions remain in-phase as a function of Zeeman field and the Majorana charge (b) freezes at its local maximum value (in red), instead of continuing the oscillation like in the non-interacting case (dashed curve).}
\label{fig4}
\end{figure*}

The effect of this peculiar evolution of the electrostatic potential has direct consequences on the 
wire spectral properties, as we analyze in Fig. \ref{fig4}, but for comparison, let us first see what happens in the non-interacting case. The wire's spectrum is shown in Fig. \ref{fig3}(a). There we can observe the emergence of low energy subgap states for $V_{Z}> \sqrt{\Delta^2+\mu^2}$, corresponding roughly to the critical field for the bulk topological transition. We also obtain the typical energy oscillations produced by overlapping Majorana wave functions due to the finite length of the wire \cite{Prada:PRB12,Das-Sarma:PRB12,Rainis:PRB13}. More insight can be obtained by analyzing the evolution of the total wire's charge 
$Q_{tot}= \int_0^L dx \left<\rho(x)\right>$ as well as
the Majorana charge $Q_M$, whose absolute value is given by
\begin{equation}
\left|Q_M\right| = \left|Q_{+1} - Q_{-1}\right|= \left| e \int_0^L dx u_L(x) u_R(x)\right|.
\label{MajoranaCharge}
\end{equation}
Here, $Q_{\pm1}$ are the charge corresponding to the even/odd lowest energy eigenstates $\psi_{\pm1}$, and $u_{L,R}$ are the electron components of the Majorana wavefunctions $\gamma_{L}=\psi_{+1}+\psi_{-1}$ and $\gamma_{R}=-i\left(\psi_{+1}-\psi_{-1}\right)$. The total charge increases in general with magnetic field but, for finite length wires, it does so by jumping abruptly a quantity equal or smaller than $e$ at each parity crossing (where the Majorana oscillations cross zero energy and the electron parity of the wire changes from even to odd or viceversa), as shown in Fig. \ref{fig2}(d), dashed curve. This abrupt change in charge is actually injected into the fermion state created by the two overlapping Majoranas and is given by $\left|Q_M\right|$ at the parity crossings. The (oscillatory) evolution of $\left|Q_M\right|$ with magnetic field is given in Fig. \ref{fig3}(b). Strikingly, $\left|Q_M\right|$ is maximum at the parity crossing, where the energy is zero, and goes to zero at the oscillation cusps.  As the length of the wire tends to infinity, $Q_M$ tends to zero (not shown). Indeed, the finite value of $Q_M$ \emph{at the parity crossings} is a direct measurement of the Majorana overlap, as shown in Ref. \cite{Dominguez:npj17}. Note that the Majorana overlap is defined similarly to the right-hand side of Eq. (\ref{MajoranaCharge}), but with the absolute value inside the integral.

The behavior of the Majorana wavefunctions is illustrated in 
Figs. \ref{fig3}(c-f). The probability density for the left and right Majorana wave functions 
exhibits an overall decay towards the center of the wire controlled by the length $\xi \sim \hbar v_F/\Delta$ and an oscillatory pattern controlled by $\lambda_F$ \cite{Kitaev:PU01,Klinovaja:PRB12}. Moreover, the number of oscillations that fit in $L$ increases by one with Zeeman field at each parity crossing.
Interestingly, we observe that the left-right oscillatory patterns are out of phase for the cases where the splitting of the MBSs is maximum, panels (c) and (e). This minimizes the left-right wave function overlap and the Majorana charge goes to zero. On the other hand, 
the oscillations are in phase (d) when the energy splitting is zero, at the parity crossings, producing a maximum in $\left|Q_M\right|$ and overlap. Although the Majorana wave functions are more strongly located at the wire edges, we note that the charge density of this fermionic state is uniform across the wire \cite{Ben-Shach:PRB15} and, thus, it is uniformly affected by the interaction with the environment when this is present.

In the presence of interactions with the image charges, the single-point parity
crossings versus $V_Z$ in the spectrum are replaced by extended regions where the subgap states remain pinned at zero energy, indicated by the red lines in Fig. \ref{fig4}(a). The abrupt jumps in $Q_{tot}$ in the non-interacting case are replaced by a linear increase along the $V_z$-values where the zero-energy pinning occurs, see Fig. \ref{fig2}(d). This is a consequence of the repulsive environment that inhibits the entrance of charge in the wire where the electron liquid behaves in an incompressible manner. On the other hand, the Majorana charge remains basically constant at the pinning plateaux, as shown in Fig. \ref{fig4}(b). The finite value of $Q_M$ in these regions indicates
that zero-energy does not imply absence of overlap between the left and right
Majorana states. This is actually a common misconception that we would like to point out here. The Majorana overlap, which is a measurement of the degree of non-locality of the two Majorana wave functions, mostly depends on the length of the nanowire (and to a lesser extent on other parameters, such as the induced superconductor gap and the Rashba coupling), but it is not necessarily correlated to the Majorana energy splitting. Different mechanisms can reduce this splitting, such as interactions with the environment as studied here, smooth potential or gap profiles \cite{Kells:PRB12,Prada:PRB12,Liu:PRB17a,Moore:17,Fleckenstein:17}, or orbital magnetic effects \cite{Dmytruk:17}, and still leave the Majorana overlap unaffected. The behavior of the Majorana wavefunctions in this case is illustrated in 
Figs. \ref{fig4}(c-f). In the pinning regions the Majorana wave functions remain practically frozen and in phase. This in turn explains why $\left|Q_M\right|$ is maximum in these regions.

The generality of these results with wire parameters is analyzed in the SI \hyperref[Section:SI4]{4}. There we show how the width of the pinning plateau evolves with $V_Z$ when we change the chemical potential, the dielectric permittivity or the width of the SC shell, and the aspect ratio of the nanowire section. We find that pinning survives for any chemical potential, while it vanishes when the attractive contribution of the SC shell becomes dominant over the dielectric repulsion.

\subsection{Effect of bulk normal leads}
\label{Section:with-leads}

\begin{figure}
\includegraphics{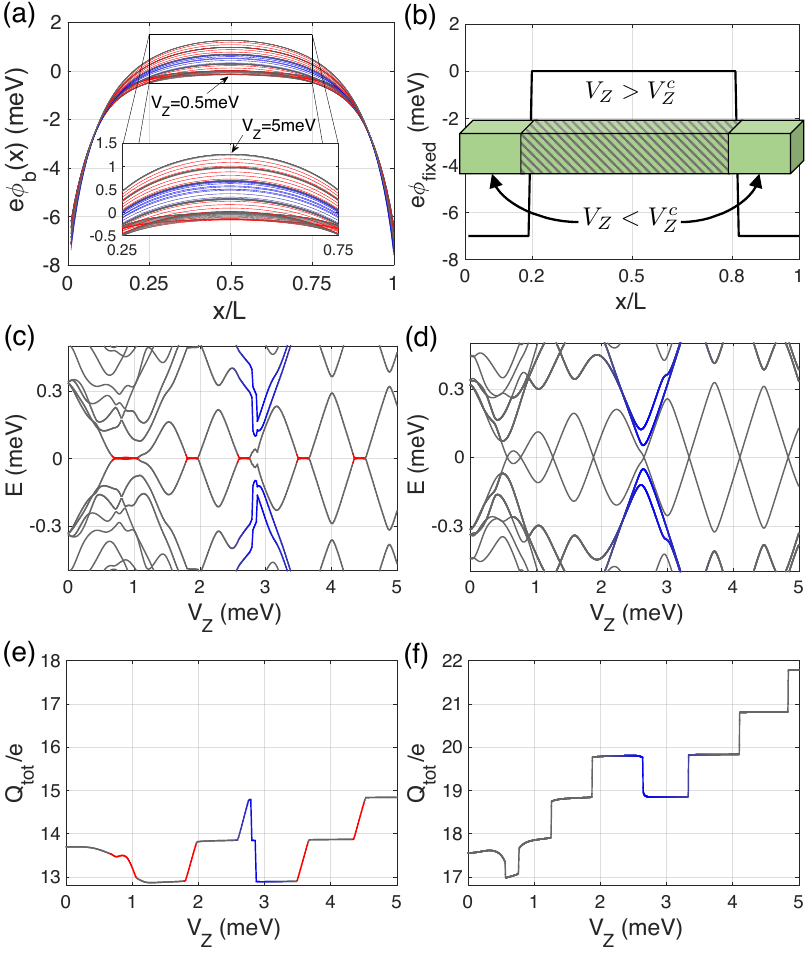}
\caption{Majorana nanowire subject to interactions from the electrostatic environment (including the influence of the bulk normal leads at its ends). (a) Self-consistent induced potential energy $e\phi_b(x)$ along the wire's length for increasing values of the Zeeman splitting. Same wire parameters as in Fig. \ref{fig2}. Note that the main effect of the bulk normal leads is to create confining potential wells at the wire edges. (b) Barrier-like potential energy profile used to mimic the self-consistent solution. Spectra of the Majorana nanowire as a function of $V_Z$ in the (c) interacting case and in the (d) non-interacting case but using the potential profile model of (b). (e,f) Evolution of the total charge $Q_{tot}$ with Zeeman splitting for the two previous cases, respectively. Red color indicates incompressible electron fluid behavior as before, while blue color indicates QD-like behavior due to the metallic contacts.}
\label{fig5}
\end{figure}

In this section we analyze the effect of including the bulk normal leads in the calculation of the induced potential $\phi_b$. Fig. \ref{fig5}(a)
illustrates the evolution of $\phi_b$ with increasing Zeeman field for the same 
choice of parameters as in the previous section but including the normal contacts.
As can be observed, while in the central region of the wire a similar repulsive 
step-like evolution with $V_z$ is found (corresponding to compressible/incompressible electron fluid behavior), significant attractive regions appear at the
wire ends produced by the metallic character ($\epsilon_M\rightarrow\infty$) of the adjacent leads. As we discuss below, these attractive regions give rise to the formation
of quantum-dot (QD) like bound states that may interact with the low energy subgap states of the Majorana wire. 

The evolution of the spectral properties and of the total charge 
$Q_{tot}$ in this case are shown in Fig. \ref{fig5}(c) and (e), respectively. On the one hand, we observe that the pinning plateaus around each parity crossing (in red) are still present although with a smaller width. On the other hand, the main effect 
of the presence of the attractive potential regions is the appearance of 
an additional four energy levels (two per contact, in blue) that approach zero energy for $V_Z$ around $2.5-3$meV. At the same time
we observe a rather abrupt decrease in the total wire charge (of roughly $2e$), see 
Fig. \ref{fig5}(e). We can associate these additional levels with QD-like
bound states arising in the attractive regions of the induced potential that anticross with the Majorana levels when their energies are on resonance \cite{Prada:PRB17,Schuray:17,Clarke:PRB17,Ptok:17}.

To demonstrate the validity of this interpretation we show in
Fig. \ref{fig5}(d) and (f) the spectral properties and the total charge evolution for an isolated wire with a simple double potential well taken to mimic the effect of the electrostatic environment, shown in (b). Notice that in this case we do not attempt a self-consistent 
calculation but rather include the Zeeman field as a rigid shift of the two
spin bands (like in the non-interacting case but with an inhomogeneous potential profile). Although the zero-energy pinning is not captured by this model, one can clearly observe the presence of
four levels coming down towards zero-energy for $V_Z$ around $2.5-3$meV, as in the 
interacting case. The presence of these states is a  consequence of the renormalization of the topological phase transition due to the electrostatic potential (either $\phi_b$ or $\phi_{fixed}$)
\begin{eqnarray}
V_{Z}^{c}=\sqrt{\left(\mu-e\phi_{b,fixed}\right)^2+\Delta^2},
\label{with-leads-critical-field}
\end{eqnarray}
which is not constant along the wire because $\phi_b$ (or $\phi_{fixed}$) depends on $x$. For the shown values of $V_Z$, only the central part of the wire is in the topological regime ($V_Z>V_Z^c$), corresponding to an effectively shorter Majorana wire, whereas the outer parts are trivial ($V_Z<V_Z^c$), corresponding to two effective QDs attached to it. Specific details of how QD-levels interact with Majorana nanowire ones can be found in Ref. \cite{Prada:PRB17,Schuray:17,Chevallier:17,Clarke:PRB17,Ptok:17}.

\begin{figure*}
\includegraphics{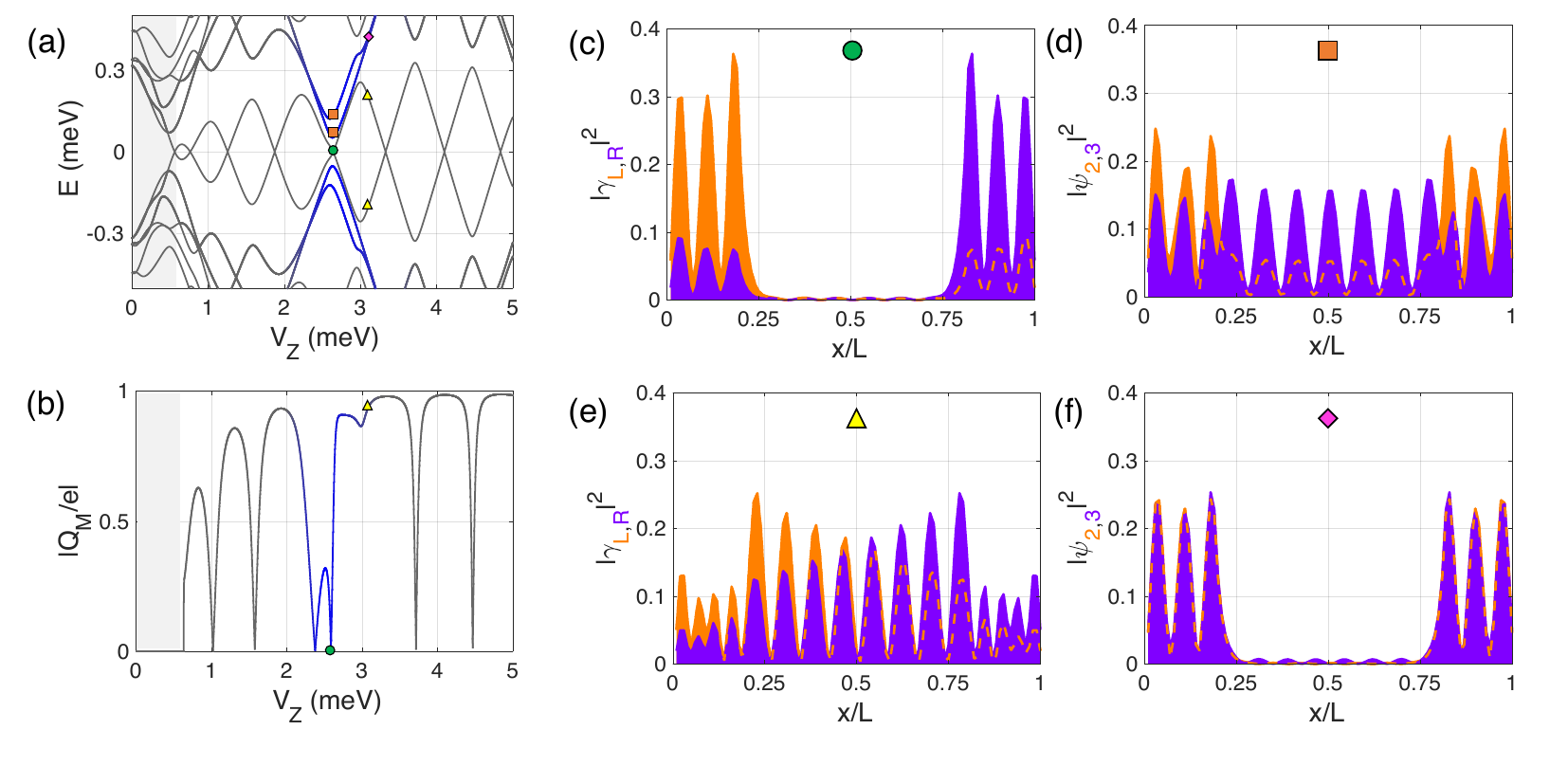}
\caption{Evolution with Zeeman field of the spectrum (a) and the absolute value of the Majorana charge $Q_M$ (b) for the barrier-like potential model of Fig. \ref{fig5}(b). Panels (c) and (e) show the wave-function probability profile (in the Majorana basis) of the two lowest energy states at the $V_Z$ values indicated in (a). Panels (d) and (f) show the same but for the second and third energy states (QD-like states). At the dot-Majorana levels anticrossing, the Majorana wave function leaks into the dot regions leaving the topological region of the wire practically void. This is manifested in $|Q_M|$ by two consecutive zeros, one per dot level (around $V_Z=2.5$meV).}
\label{fig6}
\end{figure*}

Further evidence on the nature of the low energy states at $V_Z \sim 3$meV 
is provided in Fig. \ref{fig6} where we plot the wave-function probability profiles (in the Majorana basis) of the low energy states around the QD-Majorana levels anticrossing. 
For simplicity, we consider only the case of the potential barrier model. 
At the anticrossing, the Majorana and dot states merge and cannot be really told apart, but we will refer to the two lowest in energy as Majorana levels and to the other two as dot levels. As can be observed, at the anticrossing 
the Majorana levels (green circle) leak into the QD regions leaving the central (topological) part practically void. Conversely, the two dot-like states (immediately above in energy, orange squares)
penetrate and delocalize along the wire. When the Zeeman field increases and the QD and Majorana levels are detuned, the dot states depart from low energy (pink rhombus) and from the topological part of the wire, whereas the usual overlapping behavior of the MBSs is recovered but with the Majoranas bound to the effective topological edges (yellow triangle). The absolute value of the Majorana charge vs. $V_Z$ is shown in Fig. \ref{fig6}(b), calculated considering only the two lowest energy states (as before). At the anticrossing the Majorana charge oscillation is distorted, see blue region, but the area below the curve is conserved. The missing charge in Fig. \ref{fig5} (b) does not come from the Majorana states, but from the dot ones. At the anticrossing region, the two QD states (one per potential well) that were occupied (below the Fermi level) move upwards in energy as the Zeeman field increases and cross the Fermi level, emptying themselves. This is why in the blue regions of Fig. \ref{fig5} (e) and (f) the wire's total charge does not increase at the corresponding parity crossing, but instead decreases loosing effectively twice the charge of the electron $e$. 

Finally, we would like to point out that, when the dot levels anticross the Majorana ones in a pinning region, the Majorana states detach from zero energy. This can be seen in Fig. \ref{fig5}(c) and Fig. \ref{fig1}(b). The reason is that, although in the pinning regions the Majorana energy is zero, their wave function overlap is not. It is actually maximum, as explained when discussing Fig. \ref{fig4}. Each QD acts as a local probe (one couples to the left topological region of the wire, the other to the right). If the wire's length were large (much bigger than the coherence length), left and right Majoranas would be disconnected from each other, and a local probe coupled to one of them would not be able to change its energy or perturb it. This is actually the core manifestation of their topological protection. However, when the wire's length is finite and the Majoranas overlap, each QD couples to both Majoranas at either end and their energies are modified. The typical shapes of the anticrossing were recently analyzed and can be used to quantify the degree of Majorana non-locality \cite{Prada:PRB17,Clarke:PRB17}.

\section{Conclusions}
\label{Section:Conclusions}

In this work we have studied the low energy characteristics of Majorana nanowires when including its interaction with a realistic 3D electrostatic environment. This is done by solving self-consistently the Bogoliubov-de Gennes equation together with the Poisson's equation. Typically, the total charge of the wire in equilibrium with the reservoirs increases with magnetic field (or the wire's chemical potential). However, if the electrostatic screening is smaller inside the wire than at the contacts, a repulsive interaction arises that leads to zero energy pinning around parity crossings in the wire's spectrum. While the screening due to the parent SC shell tends, in general, to reduce this pinning effect, we find that it still persists depending on the quality of the SC layer and the location of the charge density within the nanowire.
The pinning mechanism could help explain the precise shape of the Majorana oscillations (or lack thereof) observed in some dI/dV experiments, which exhibit substantial deviations from the predictions of simple models for finite length wires.

On the other hand, and more importantly, the self consistent solution of the electrostatic potential varies non-homogeneously along the wire. It is relatively flat in the central region but, due to the screening from the left/right metallic contacts, it becomes strongly negative at the edges. This creates potential wells that confine QD-like states at the ends of the wire, which appear in the spectrum as discrete states within the induced gap that disperse with Zeeman energy or chemical potential. These QD levels interact with the Majorana states in a specific way which is strongly dependent on the Majorana wavefunction, and particularly on its degree of spatial non-locality. The pinning mechanism and the coupling to QD-like states compete against each other, so that the pinned zero-energy plateaux may become lifted at resonance with the dot states, thus revealing their electrostatic origin (as opposed to true wavefunction non-locality).

%%%%%%%%%%%%%%%%%%%%%%%%%%%%%%%%%%%%%%%%%%%%%%%%%%%%%%%%%%%%%%%%%%%%%
%% When referencing objects, LaTeX offers a label--ref mechanism.
%% The beilstein class extends this approach with the \cref command,
%% that adds the corresponding type of the object as well.
%
%
%% Tables, figures and schemes must have a single column or double
%% column width. To make life easier, some commands are defined.
%%
%% Captions (legends) will always be added at the correct place no
%% matter where you put in the source code.
%% Please note: labels always have to come /after/ the \caption.
%%%%%%%%%%%%%%%%%%%%%%%%%%%%%%%%%%%%%%%%%%%%%%%%%%%%%%%%%%%%%%%%%%%%%

%%%%%%%%%%%%%%%%%%%%%%%%%%%%%%%%%%%%%%%%%%%%%%%%%%%%%%%%%%%%%%%%%%%%%
%% The "Acknowledgements" section can be given in all manuscripts.
%% This should be done within the ``acknowledgements'' environment,
%% which will make the correct section title.
%%%%%%%%%%%%%%%%%%%%%%%%%%%%%%%%%%%%%%%%%%%%%%%%%%%%%%%%%%%%%%%%%%%%%
\begin{acknowledgements}
We thank P. San-Jose and R. Aguado for valuable discussions. Research supported by the Spanish Ministry of Economy and Competitiveness through Grants FIS2014-55486-P, FIS2016-80434-P (AEI/FEDER, EU), the Ram\'on y Cajal programme Grants RYC-2011-09345 and the Mar\'ia de Maeztu Programme for Units of Excellence in R\&D (MDM-2014-0377).  
\end{acknowledgements}

\appendix
\section*{Supporting information}
\section*{SI 1: Expressions for the induced potential using the image charge method}
\label{Section:SI1}
\subsection*{Introduction}
The electrostatic potential $\phi_b$ produced by the environment is calculated from the interaction between the nanowire charge density and bound charges it creates at the surrounding
medium shown in Fig. (\ref{fig1}) of the main text. To do this we use the method of the image charges. We model the nanowire as a semiconductor rod of square section of relative permittivity $\epsilon$, length $L$ and rectangular
section of width $W=2R$, being $R$ the half-width. We assume that the charge density $\rho\left(\vec{r}\right)$ of the nanowire is located along its symmetry axis ($x$-axis) as a linear charge density. The nanowire faces are in contact with different dielectric materials
of permittivities $\epsilon_{1}$ (substrate), $\epsilon_{2}$ (superconducting shell), $\epsilon_{3}$
and $\epsilon_{4}$ (surrounding medium); while two metal leads of permittivities $\epsilon_{M_{1}}$
and $\epsilon_{M_{2}}$ are placed at both ends of the nanowire. In
order to give more insight on the solution of this problem, we solve first some simpler cases. In the first section we obtain the electrostatic potential created by a linear charge density placed before one, between two, and before two infinite planes. These problems can be 
understood as 1D problems. In the second section we obtain the potential with all four surrounding media but without the bulk leads, which can be treated as a 2D problem. Finally, in the third section we obtain the full model including also the interaction with the leads.

\subsection*{Section I: One dimension}

\subsubsection*{A. One infinite plane}

\begin{figure}
\includegraphics{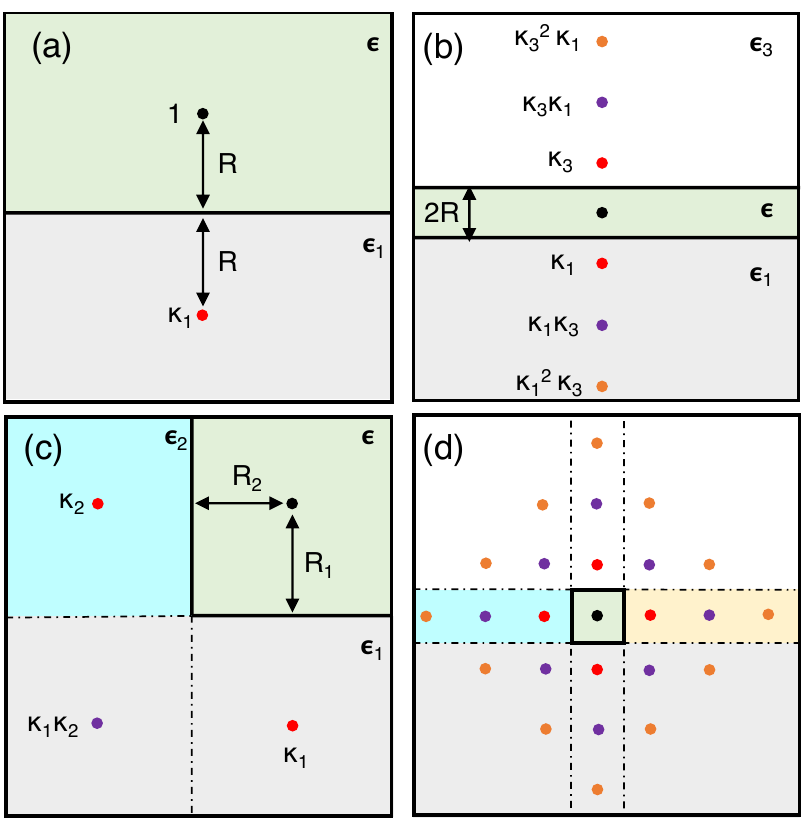}
\caption{Configuration of the image charges produced by the original charge (in black) studied in different Sections. The original charge is in a medium of dielectric constant $\epsilon$ (green color in the plots) that extends in the third direction. (a) Sec. IA: a charge in front of an infinite semi-space of permittivity $\epsilon_{1}$ at a distance $R$. (b) Sec. IB: a charge between two semi-infinite and parallel regions of permittivities $\epsilon_{1}$ and $\epsilon_{3}$. (c) Sec. IIA: a charge close to the intersection between two media of permittivities $\epsilon_{1}$ and $\epsilon_{2}$. (d) Sec. IIB: a charge between four different media. As before, the black point depicts the real charge, while the red, purple and orange points depict the first, second and third image charges found in the first, second and third steps of the image method procedure, respectively. For (a) and (c) all the image charges are shown, whereas for (b) and (d) there is an infinite number of them and only the first ones are shown. Black solid lines are interfaces, and black dashed lines are image interfaces.}
\label{figSI1}
\end{figure}

The solution of this problem can be found in
textbooks on electromagnetism \cite{Reitz:Book97}. When  one charge $q$ is placed (at the origin of coordinates) in a medium of dielectric constant $\epsilon$ and at a distance $R$ from the interface
between this and another medium of permittivity $\epsilon_{1}$, a bound charge of magnitude $\kappa_1q$ appears spread at the interface, where
\begin{equation}
\kappa_{1}\equiv\frac{\epsilon-\epsilon_{1}}{\epsilon+\epsilon_{1}}.
\end{equation}
The effect of this bound charge is equivalent to that of a point charge of the same magnitude and located at a specular distance with respect to the plane from the original charge, see Fig. \ref{figSI1}(a). This is why it is called an \emph{image} charge.
The classical electrostatic potential due to the interaction between the original and the image charge through the Coulomb's law takes the form
\begin{equation}
\phi_{b}\left(x\right)=\frac{1}{4\pi\epsilon\epsilon_{0}}\frac{\kappa_{1}q}{\sqrt{\left(2R\right)^{2}+x^{2}}},
\label{eqPhibIA}
\end{equation}
where $\epsilon_{0}$ is the vacuum permittivity.
Because $\phi_{b}$ is linear in $q$, this result can be generalized to an arbitrary 1D density charge $\rho\left(x\right)$:
\begin{equation}
\phi_{b}\left(x\right)=\frac{1}{4\pi\epsilon\epsilon_{0}}\int\frac{\kappa_{1}\rho\left(x'\right)}{\sqrt{\left(2R\right)^{2}+\left(x-x'\right)^{2}}}dx'.
\end{equation}

As argued in Ref. \cite{Dominguez:npj17}, since the bound charges are distinguishable from the nanowire
charges (cannot tunnel in and out), this potential can be
directly transformed into a quantum operator as a Hartree interaction
without any Fock correction. Assuming a purely local polarizability
(Thomas-Fermi limit), the density operator may be transformed as $\rho\left(x'\right)\rightarrow\left\langle \rho\left(x'\right)\right\rangle $,
which is perfectly equivalent to the above classical equation. Then, the potential takes the form
\begin{equation}
\phi_{b}\left(x\right)=\int V_{b}\left(x,x'\right)\left\langle \hat{\rho}\left(x'\right)\right\rangle dx',
\label{potential}
\end{equation}

\noindent where the kernel of the interaction $V_{b}\left(x,x'\right)\equiv\kappa_{1}/4\pi\epsilon\epsilon_{0}\sqrt{\left(2R\right)^{2}+\left(x-x'\right)^{2}}$
encodes the geometrical information of the interaction. 

We note that Eq. (\ref{potential}) is general for any geometry and charge density. For this reason, in the following sections we first obtain the kernel function $V_{b}$ for a given geometry using a single original charge, and then we generalize our results to an arbitrary density $\rho\left(x\right)$ and to its corresponding quantum expression using Eq. (\ref{potential}). 

\subsubsection*{B. Two opposite infinite planes}

We now consider a charge between two infinite and parallel planes separating the nanowire's dielectric constant $\epsilon$ from two other media of permittivities $\epsilon_{1}$ and
$\epsilon_{3}$ [see Fig. \ref{figSI1}(b)]. In order to try to satisfy the boundary conditions imposed by the Gauss' law, in a first step two image charges $\kappa_{1}q$ and $\kappa_{3}q$
are required in each dielectric medium at the same distance $R$
from the interface [see red dots in Fig. \ref{figSI1}(b)]. However, each image charge does not satisfy the
boundary condition with respect to the opposite interface. For this reason, in a second step, two
(accidentally equals) additional image charges of magnitude $\kappa_{1}\kappa_{3}q$
are required at a distance $3R$ from each interface [see purple dots in Fig. \ref{figSI1}(b)]. But, again, these additional image charges
do not satisfy the boundary conditions with respect to the opposite interfaces and another step must be taken.
It is possible to see that for each image charge $q_{\alpha}^{\left(n\right)}$ in the dielectric $\alpha$ created at the $n$-th step of the image charge method, another image charge
\begin{equation}
q_{\beta}^{\left(n+1\right)}=\kappa_{\beta}q_{\alpha}^{\left(n\right)}
\end{equation}
is required in the opposite $\beta$ dielectric at a distance $2\left(n+1\right)R$
from the original charge $q$ in order to satisfy the boundary conditions, with $q_{\alpha,\beta}^{\left(0\right)}=1$.
Thus, the interaction kernel takes the form of an infinite series
\begin{equation}
V_{b}\left(x\right)=\frac{1}{4\pi\epsilon\epsilon_{0}}\sum_{n=1}^{\infty}\left(\frac{q_{1}^{\left(n\right)}+q_{3}^{\left(n\right)}}{\sqrt{x^{2}+\left(2nR\right)^{2}}}\right).
\end{equation}
Note that each term of the sum decreases with $n$ as $V_{b}^{\left(n\right)}\sim\kappa^{n}/n$. Since $\left|\kappa\right|<1$ for a dielectric medium and $\kappa_{M}=-1$ for a metallic one (where $\epsilon_{M}\rightarrow\infty$), then the kernel converges to $V_{b}\sim\ln\left(1-\kappa\right)$ in spite of the infinite summation.

\subsubsection*{C. Two consecutive infinite planes}

\begin{figure}
\includegraphics{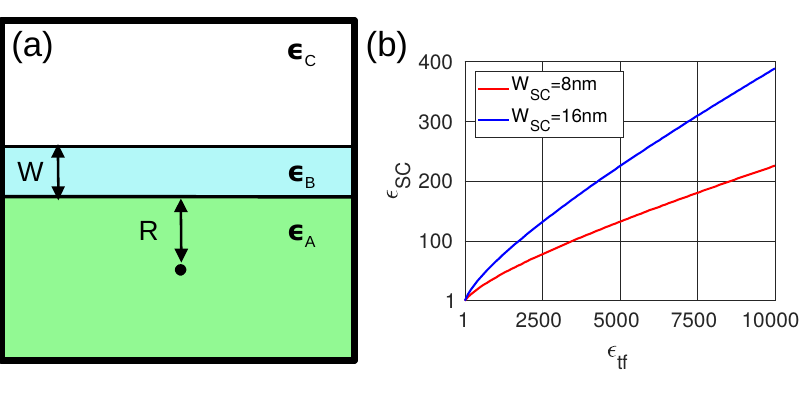}
\caption{(a) System studied in Sec. IC. A charge (black dot) is in a medium of dielectric constant $\epsilon_A$ (the nanowire) and placed at a distance $R$ from a thin material of permittivity $\epsilon_B$ and thickness $W$ (the SC shell). Above the thin film there is another semi-infinite space of permittivity $\epsilon_C$ (the vacuum). When all the permittivities are finite, bound charges arise in both surrounding mediums. (b) Effective SC permittivity $\epsilon_{\rm{SC}}$ calculated using Eq. \ref{epsilon_eff} vs the SC thin film permittivity $\epsilon_{\rm{tf}}$ for two different film thicknesses $W_{SC}$. Parameters are the same as in the main text.}
\label{figSI4}
\end{figure}

Finally, we consider the case depicted in Fig. \ref{figSI4}(a) where a (real) charge $q$ is placed in a dielectric material characterized by a permittivity $\epsilon_A$. This charge is at a distance $R$ from a first interface with a material of dielectric constant $\epsilon_B$ and width $W$, and at a distance $R+W$ from a second (parallel) interface with another medium of permittivity $\epsilon_C$. As we know, the real charge $q$ creates an image charge $\kappa_B q$ at a distance $2R$ from it inside the $\epsilon_B$ medium. However, the potential created by both charges is only valid inside the $\epsilon_A$ medium. The potential created inside the $\epsilon_B$ medium can be found using the image method \cite{Reitz:Book97}: it is the same potential than that created by an image charge of magnitude $q[2\epsilon_A/(\epsilon_A+\epsilon_B)]$ located at the same position of the real charge. From this point, the same steps explained in the previous subsection can be followed, and once the infinite series is obtained, the potential can be transformed back to the $\epsilon_A$ medium. Thus, one can prove that the bound charges potential is given by
\begin{multline}
V_{b}=\frac{q}{4\pi\epsilon_A\epsilon_0}\left(\frac{1}{\epsilon_A+\epsilon_B}\right)\left[\frac{\epsilon_A-\epsilon_B}{\sqrt{x^2+(2R)^2}}\right.+ \\
+\left.\frac{4\epsilon_A\epsilon_B}{\epsilon_B-\epsilon_A}\sum_{n=1}^{\infty}{\frac{\left(\frac{(\epsilon_B-\epsilon_A)(\epsilon_B-\epsilon_C)}{(\epsilon_B+\epsilon_A)(\epsilon_B+\epsilon_C)}\right)^n}{\sqrt{x^2+(2R+2nW)^2}}}\right].
\end{multline}
Since this expression is rather complex, it is convenient to replace the effect of both media $\epsilon_B$ and $\epsilon_C$ by just one characterized by an effective permittivity $\epsilon_{\rm{eff}}$ which, from an electrostatic point of view, is equivalent. Hence, the bound charges potential would be given by Eq. \ref{eqPhibIA} with $\epsilon\rightarrow\epsilon_A$ and $\epsilon_1\rightarrow\epsilon_{\rm{eff}}$. Comparing both equations, the effective permittivity (at $x=0$) is
\begin{equation}
\epsilon_{\rm{eff}}=\epsilon_A\frac{1-\kappa_{\rm{eff}}}{1+\kappa_{\rm{eff}}},
\label{epsilon_eff}
\end{equation}
where
\begin{multline}
\kappa_{\rm{eff}}=\frac{1}{\epsilon_A+\epsilon_B}\left[\epsilon_A-\epsilon_B+\right. \\
\left.+\frac{4\epsilon_A\epsilon_B}{\epsilon_B-\epsilon_A}\sum_{n=1}^{\infty}{\frac{1}{1+n\frac{W}{R}}\left(\frac{(\epsilon_B-\epsilon_A)(\epsilon_B-\epsilon_A)}{(\epsilon_B+\epsilon_C)(\epsilon_B+\epsilon_C)}\right)^n}\right].
\end{multline}
We note that this system corresponds to the nanowire-SC shell-vacuum double interface of Fig. \ref{fig1}(a). There, the effect of the shell finite width and the vacuum on top has been condensed in an effective SC permittivity. This means that, in Eq. \ref{epsilon_eff}, $\epsilon_{\rm{eff}}\rightarrow\epsilon_{\rm{SC}}$, $\epsilon_{A}\rightarrow\epsilon$, $\epsilon_{C}\rightarrow\epsilon_a$ and $\epsilon_{B}$ is the true SC thin film permittivity $\epsilon_{\rm{tf}}$.

In Fig. \ref{figSI4}(b) we show the effective SC permittivity $\epsilon_{\rm{SC}}$ considered in the main text as a function of $\epsilon_{\rm{tf}}$ for two different shell widths $W_{SC}$. Notice that, as the SC shell becomes thinner (red curve corresponds to 8nm), the effective permittivity gets more renormalized. A value $\epsilon_{\rm{SC}}\sim 100$ corresponds to $\epsilon_{\rm{tf}}\sim 4000$.

\subsection*{Section II: Two dimensions}

\subsubsection*{A. One rectangular corner}

This is also a textbook problem \cite{Reitz:Book97}. A charge $q$ inside a dielectric with permittivity $\epsilon$ is placed
at a distance $R_{1}$ from dielectric $\epsilon_{1}$ and at a different
distance $R_{2}$ form dielectric $\epsilon_{2}$, which are perpendicular
to one another [see Fig. \ref{figSI1}(c)].  To satisfy the boundary conditions, two image charges $\kappa_{1}q$ and $\kappa_{2}q$
are required in each dielectric at $\left(x,-2R_{1},0\right)$ and
$\left(x,0,-2R_{2}\right)$, see red dots. Because of these, another 
image charge of magnitude $\kappa_{1}\kappa_{2}q$ is required at $\left(x,-2R_{1},-2R_{2}\right)$, purple dot.
In this case, and due to the closed geometry of the problem, three image charges are enough to satisfy the boundary conditions.
The kernel function takes thus the form
\begin{multline}
V_{b}\left(x\right)=\frac{1}{4\pi\epsilon\epsilon_{0}}\left(\frac{\kappa_{1}}{\sqrt{x^{2}+\left(2R_{1}\right)^{2}}}+\frac{\kappa_{2}}{\sqrt{x^{2}+\left(2R_{2}\right)^{2}}}\right.+\\
\left.+\frac{\kappa_{1}\kappa_{2}}{\sqrt{x^{2}+\left(2R_{1}\right)^{2}+\left(2R_{2}\right)^{2}}}\right).
\end{multline}

\subsubsection*{B. Four rectangular corners: interaction without the leads}

Now a charge $q$ is placed inside an infinite wire of rectangular section and of permittivity
$\epsilon$, see Fig. \ref{figSI1}(d). This charge is at a distance $R_{1}$ from two parallel flat dielectrics
with permittivities $\epsilon_{1}$ and $\epsilon_{3}$, and at a
distance $R_{2}$ from another two parallel dielectrics with
permittivities $\epsilon_{2}$ and $\epsilon_{4}$ which are perpendicular
to the previous ones. Combining what we have learned in the previous Sections, to satisfy the boundary conditions an infinite ensemble of image charges have to be placed in the different dielectrics as shown in Fig. \ref{figSI1}(d). For each image charge in
one dielectric, another one appears at a specular distance from the opposite interface, while for each two image charges placed in two perpendicular media, just
one more appears at the corner. The electrostatic potential due to all these charge is given by
\begin{multline}
V_{b}\left(x\right)=\frac{1}{4\pi\epsilon\epsilon_{0}}\left[\sum_{n,m=1}^{\infty}\left(\frac{\left(q_{1}^{\left(n\right)}+q_{3}^{\left(n\right)}\right)\left(q_{2}^{\left(m\right)}+q_{4}^{\left(m\right)}\right)}{\sqrt{x^{2}+\left(2nR_{1}\right)^{2}+\left(2mR_{2}\right)^{2}}}\right)+\right.\\
\left.+\sum_{n=1}^{\infty}\left(\frac{q_{1}^{\left(n\right)}+q_{3}^{\left(n\right)}}{\sqrt{x^{2}+\left(2nR_{1}\right)^{2}}}+\frac{q_{2}^{\left(n\right)}+q_{4}^{\left(n\right)}}{\sqrt{x^{2}+\left(2nR_{2}\right)^{2}}}\right)\right],
\end{multline}

\noindent where:
\begin{equation}
\begin{cases}
\begin{array}{ccc}
q_{1}^{\left(n+1\right)}=\kappa_{1}q_{3}^{\left(n\right)}, &  & q_{2}^{\left(n+1\right)}=\kappa_{2}q_{4}^{\left(n\right)},\end{array}\\
\begin{array}{ccc}
q_{3}^{\left(n+1\right)}=\kappa_{3}q_{1}^{\left(n\right)}, &  & q_{4}^{\left(n+1\right)}=\kappa_{4}q_{2}^{\left(n\right)},\end{array}\\
q_{\alpha}^{\left(0\right)}=1\;\;\forall\alpha=\left\{ 1,2,3,4\right\} .
\end{cases}
\end{equation}

If the wire's section is square ($R_{1}=R_{2}$), then the kernel function can be
rewritten in a more compact way:
\begin{widetext} 

\begin{equation}
V_{b}\left(x\right)=\frac{1}{4\pi\epsilon\epsilon_{0}}\sum_{n,m=0}^{\infty}\left(\frac{\left(q_{1}^{\left(n\right)}+q_{3}^{\left(n\right)}-\delta_{n,0}\right)\left(q_{2}^{\left(m\right)}+q_{4}^{\left(m\right)}-\delta_{m,0}\right)}{\sqrt{x^{2}+\left(2nR\right)^{2}+\left(2mR\right)^{2}}}\right)\left(1-\delta_{n+m,0}\right).
\end{equation}

\end{widetext}
Note that the number of charges increases as $4n$ at each $n$-th step of the image charge method, whereas the rest of the expression inside the brackets decreases as $\kappa^{n}/n$ as before. Thus, each term of the sum goes as $V_{b}^{(n)}\sim\kappa^{n}$ and the infinite sum is proportional to $V_{b}\sim\kappa/\left(1-\kappa\right)$ if $\left|\kappa\right|<1$, so the convergence of the kernel is ensured in this case as well.

\subsection*{Section III: The full-model}

Finally, we solve the full system of Fig. (1) of the main text. Now, apart from the four dielectric media in each of the four faces of the square section, there are another
two faces in the $x$-direction in contact to metallic regions. We consider that the nanowire has a square section of semi-width $R$. First, we assume that the charge $q$ is placed at
the coordinates origin and at the same distance $R$ from each metallic region $M_{1}$ and $M_{2}$. Following the same procedure as before we obtain
\begin{widetext} 
\begin{multline}
V_{b}\left(x\right)=\frac{1}{4\pi\epsilon\epsilon_{0}}\sum_{n,m,k=0}^{\infty}\left(\frac{\left(q_{1}^{\left(n\right)}+q_{3}^{\left(n\right)}-\delta_{n,0}\right)\left(q_{2}^{\left(m\right)}+q_{4}^{\left(m\right)}-\delta_{m,0}\right)q_{M_{1}}^{\left(k\right)}}{\sqrt{\left(x+2kR\right)^{2}+\left(2nR\right)^{2}+\left(2mR\right)^{2}}}+\right.\\
\left.+\frac{\left(q_{1}^{\left(n\right)}+q_{3}^{\left(n\right)}-\delta_{n,0}\right)\left(q_{2}^{\left(m\right)}+q_{4}^{\left(m\right)}-\delta_{m,0}\right)\left(q_{M_{2}}^{\left(k\right)}-\delta_{k,0}\right)}{\sqrt{\left(x-2kR\right)^{2}+\left(2nR\right)^{2}+\left(2mR\right)^{2}}}\right)\left(1-\delta_{n+m+k,0}\right),\label{1-1}
\end{multline}
\end{widetext}
where
\begin{equation}
\begin{cases}
\begin{array}{ccc}
q_{M_{1}}^{\left(n+1\right)}=\kappa_{M_{1}}q_{M_{2}}^{\left(n\right)}, &  & q_{M_{2}}^{\left(n+1\right)}=\kappa_{M_{2}}q_{M_{1}}^{\left(n\right)},\end{array}\\
q_{\alpha}^{\left(0\right)}=1\;\;\forall\alpha=\left\{ M_{1},M_{2}\right\} .
\end{cases}
\end{equation}

If now the charge $q$ is placed at an arbitrary position $x'$ inside
the nanowire, and the metal $M_{1}$ interface is at $x=0$
and the $M_{2}$ interface is  at $x=L$, then the kernel function is given by

\begin{widetext} 
\begin{multline}
V_{b}\left(x\right)=\frac{1}{4\pi\epsilon\epsilon_{0}}\sum_{n,m,k=0}^{\infty}\left(\frac{\left(q_{1}^{\left(n\right)}+q_{3}^{\left(n\right)}-\delta_{n,0}\right)\left(q_{2}^{\left(m\right)}+q_{4}^{\left(m\right)}-\delta_{m,0}\right)q_{M_{1}}^{\left(k\right)}}{\sqrt{\left(x-\left(-1\right)^{k}\left(2^{\mathrm{floor}\left(\frac{k}{2}+1\right)}L-2L+x'\right)\right)^{2}+\left(2nR\right)^{2}+\left(2mR\right)^{2}}}+\right.\\
\left.+\frac{\left(q_{1}^{\left(n\right)}+q_{3}^{\left(n\right)}-\delta_{n,0}\right)\left(q_{2}^{\left(m\right)}+q_{4}^{\left(m\right)}-\delta_{m,0}\right)\left(q_{M_{2}}^{\left(k\right)}-\delta_{k,0}\right)}{\sqrt{\left(x+\left(-1\right)^{k}\left(2^{\mathrm{floor}\left(\frac{k+1}{2}\right)}L-x'\right)\right)^{2}+\left(2nR\right)^{2}+\left(2mR\right)^{2}}}\right)\left(1-\delta_{n+m+k,0}\right).\label{Full_potential}
\end{multline}
\end{widetext}

If $L$ is large enough compared to $2R$, we can take into account only the lowest order of the image charges at the metals, $q_{M_{i}}^{(k=0,1)}$, and the number of charges at each $n$-th step of the image charge method increases only as $\sim12n$. Then, the system follows the same convergence criterion as in the previous Section. If $L\sim2R$, Eq. (\ref{Full_potential}) converges as well, but the demonstration is longer.
%If this is not true ($L\sim2R$), then the
%number of charges increase at each step as $\sim6n^{2}$
%(cubic area). Since the rest of expression within the brackets goes like $\sim\kappa^{n}/n$ then the %infinite sum of the kernel is proportional to $V_{b}\sim\kappa/\left(1-\kappa\right)^2$, so that the kernel %convergence is also ensured.

%%%%%%%%%%%%%%%%%%%SI2
\section*{SI 2: Further details on numerical methods}
\label{Section:SI2}
\subsection*{Mean field approximation to treat electron-electron interactions}
We want to solve the energy spectrum of the nanowire Hamiltonian $\hat{H}$ when the interaction between electrons $\hat{\phi}$ is included. In general, this interaction can be written in second quantization as
\begin{equation}
\hat{\phi}=\sum_{\alpha,\beta}\check{c}_{\alpha}^{\dagger}\check{c}_{\alpha}V_{\alpha\beta}\check{c}_{\beta}^{\dagger}\check{c}_{\beta},\label{1}
\end{equation}
where $\check{c}_{\alpha}^{\dagger}$, $\check{c}_{\alpha}$ are defined as the \emph{Nambufied} vector of operators
\begin{equation}
\check{c}^{\dagger}_{\alpha}=\left(c_{1\uparrow}^{\dagger},c_{1\downarrow}^{\dagger},c_{2\uparrow}^{\dagger},\mathrm{...},c_{N\downarrow}^{\dagger},c_{1\uparrow},c_{1\downarrow},c_{2\uparrow},\mathrm{...},c_{N\downarrow}\right).
\end{equation}
Here, $c_{i\sigma}^{\dagger}$ and $c_{i\sigma}$ are
electron creator/annihilation operators with quantum numbers $\alpha$. $V_{\alpha\beta}$ above encodes the electronic interaction. Therefore, Greek indexes $\alpha, \beta$ encode both particle/hole character, spin ($\uparrow,\downarrow$), and any other indexes the electron might have, such as site index $i=1,...,N$ in a tight-binding description.

To treat the quartic interaction we resort to a mean field approach called the Hartree-Fock-Bogoliubov
(HFB) approximation. Using the Wick's theorem and neglecting fluctuations and constant terms we can write

\[
\hat{\phi}_{\rm{eff}}=\sum_{\alpha,\beta}V_{\alpha\beta}\left[\left\langle \check{c}_{\alpha}^{\dagger}\check{c}_{\alpha}\right\rangle \check{c}_{\beta}^{\dagger}\check{c}_{\beta}+\left\langle \check{c}_{\beta}^{\dagger}\check{c}_{\beta}\right\rangle \check{c}_{\alpha}^{\dagger}\check{c}_{\alpha}+\left\langle \check{c}_{\alpha}^{\dagger}\check{c}_{\beta}\right\rangle \check{c}_{\beta}^{\dagger}\check{c}_{\alpha}+\right.
\]

\begin{equation}
\left.+\left\langle \check{c}_{\alpha}\check{c}_{\beta}^{\dagger}\right\rangle \check{c}_{\alpha}^{\dagger}\check{c}_{\beta}-\left\langle \check{c}_{\alpha}^{\dagger}\check{c}_{\beta}^{\dagger}\right\rangle \check{c}_{\alpha}\check{c}_{\beta}-\left\langle \check{c}_{\alpha}\check{c}_{\beta}\right\rangle \check{c}_{\alpha}^{\dagger}\check{c}_{\beta}^{\dagger}\right].\label{2}
\end{equation}

\noindent The first two terms are known as Hartree
terms, which include information about direct (repulsive/attractive) interaction
between electrons. The second and third are the Fock terms, which
include the exchange interaction due to the electron indistinguishable properties.
These terms ensure that non-physical self-interactions introduced by
the first terms are cancelled. The last two are known
as Bogoliubov terms, which include possible pairing correlations between
electrons.

We want to rewrite Eq. (\ref{2}) in a more compact manner. In order
to do that, we define two matrices:
\begin{itemize}
\item The lambda matrix:
\begin{equation}
\Lambda\equiv\mathbb{I}_{space}\otimes\mathbb{I}_{2\times2}\otimes\sigma_{x},
\end{equation}
\noindent where $\mathbb{I}_{space}$ is the identity matrix in 
real space (for a one-dimensional tight-binding model with $N$ sites, this matrix is the $N\times N$ identity), $\mathbb{I}_{2\times2}$ is
the identity matrix in spin space, and $\sigma_{x}$ is the Pauli
$x$-matrix in Nambu space. This matrix satisfies the property $\check{c}^{\dagger}=\Lambda\check{c}$. 
\end{itemize}
\begin{itemize}
\item The density matrix:
\begin{equation}
\rho_{\alpha\beta}\equiv\left\langle \check{c}_{\alpha}^{\dagger}\check{c}_{\beta}\right\rangle .
\end{equation}
We can express this matrix in terms of the eigenvectors $\gamma_{n}$ of the Hamiltonian
$\hat{H}+e\hat{\phi}_{eff}$, which are related to the $\check{c}_{\alpha}$'s through a unitary transformation $\Psi$ as $\gamma_{n}=\Psi_{n\alpha}\check{c}_{\alpha}$. Then
\begin{equation}
\rho_{\alpha\beta}=\left\langle \check{c}_{\alpha}^{\dagger}\check{c}_{\beta}\right\rangle =\Psi_{\alpha n}^{*}\left\langle \gamma_{n}^{\dagger}\gamma_{m}\right\rangle \Psi_{m\beta}=\left(\Psi^{\dagger}F\Psi\right)_{\alpha\beta}, \label{4}
\end{equation}

\noindent where $F_{nm}\equiv\left\langle \gamma_{n}^{\dagger}\gamma_{m}\right\rangle =f_{FD}\left(\epsilon_{n}\right)\delta_{nm}$
is the Fermi-Dirac distribution matrix. 
\end{itemize}

Using these two matrices, Eq. (\ref{2}) can be rewritten as
\begin{multline}
\phi_{\rm{eff}}=2\mathcal{D}\left[V\cdot d\left\{ \rho\right\} \right]+\Lambda\cdot\left(V\star\rho\right)\cdot\Lambda+V\star\left(\Lambda\cdot\rho\cdot\Lambda\right)-\\
-\Lambda\cdot\left(V\star\left(\rho\cdot\Lambda\right)\right)-\left(V\star\left(\Lambda\cdot\rho\right)\right)\cdot\Lambda,\label{3}
\end{multline}

\noindent where we assume a symmetric potential $V_{\alpha\beta}=V_{\beta\alpha}$. Here we have used the notation
 $\mathcal{D}\left[v\right]$ as the diagonal matrix
with vector $v$ in its diagonal, $d\left\{ A\right\} $ as a column
vector whose elements are the diagonal elements of matrix $A$, the
dot product $A\cdot B$ as a matrix product, and the star product
$A\star B$ as an element wise product between matrices.

However, Eq. (\ref{3}) does not have Nambu structure because in general $V_{i\sigma\tau,i\sigma\tau}\neq V_{i\sigma\bar{\tau},i\sigma\bar{\tau}}$, so Bogoliubov-de
Gennes formalism cannot be applied. We symmetrize this expression by doing
\begin{equation}
\check{c}^{\dagger}\phi_{\rm{eff}}\check{c}=\check{c}^{\dagger}\left(\frac{\phi_{\rm{eff}}-\Lambda \phi_{\rm{eff}}^{t}\Lambda}{2}\right)\check{c} + \rm{cnst.},
\end{equation}

\noindent where we have used the anticommutation relation $\left\{ c,c^{\dagger}\right\} =1$, the property $\Lambda^{t}=\Lambda$ and we neglect 
constant terms again. Thus, the general interaction between electrons in the HFB approximation
can be written as
\begin{equation}
\phi^{\rm{HFB}}=\frac{1}{2}\check{c}^{\dagger}\left(\phi_{\rm{eff}}-\Lambda \phi_{\rm{eff}}^{t}\Lambda\right)\check{c},\label{5}
\end{equation}

\noindent where $\phi_{\rm{eff}}$ is given by Eq. (\ref{3}).

\subsection*{Inclusion of the intrinsic interaction}

The interaction between the electrons inside the nanowire (\emph{intrinsic} interaction) is given, in principle, by the bare Coulomb potential in one dimension. Taking into account the finite radius (half-width) $R$ of the wire, a more precise form for the interaction is \cite{Giuliani:05}
\begin{equation}
\label{Giuliani}
V\left(x\right)=\frac{\sqrt{\pi}}{4\pi\epsilon\epsilon_{0}R}e^{x^{2}/R^{2}}\mathrm{Erfc}\left(\frac{\left|x\right|}{R}\right),
\end{equation}
where $x$ is the distance between electrons.

When we consider this potential only at the Hartree level, we find zero energy pinning around parity crossings, just like we did with the extrinsic interaction. However, this pinning is unphysical since it comes from spurious self-interaction terms introduced by the Hartree approximation \cite{Dominguez:npj17}. For the intrinsic interaction it is thus necessary to include the Fock correction due to the indistinguishability of electrons in the nanowire.

If we consider the bare interaction, Eq. (\ref{Giuliani}), in the Fock terms, an overcompensation of the pinning effect is found and unphysical jumps appear at each parity crossing. To cure this problem, we introduce screening in the quasi-static Thomas-Fermi limit so that the potential in the Fock terms acquires an additional exponential decay $e^{-\left|x\right|/\lambda_{TF}}$ that depends on the Thomas-Fermi length $\lambda_{TF}$ (which should be of the order of the Fermi wavelength). In this case, we find that the parity crossings induced by the intrinsic interaction are suppressed, in agreement with the self-interaction argument.

If the nanowire is discretized using a thigh-binding model, the interaction can be written as
\begin{multline}
V_{\alpha,\beta}=\frac{\sqrt{\pi}}{4\pi\epsilon\epsilon_{0}R}\exp\left\lbrace\left(\frac{i-j}{aR}\right)^{2}-\frac{\left|i-j\right|}{a\lambda_{TF}}\right\rbrace\cdot\\
\cdot\mathrm{Erfc}\left(\frac{\left|i-j\right|}{aR}\right)\left[1-\delta_{\alpha,\beta}\right],
\end{multline}
where the indexes $\alpha=\left\lbrace i,\sigma,\tau\right\rbrace$ and $\beta=\left\lbrace j,\sigma',\tau'\right\rbrace$ include all the quantum numbers of the electrons, $a$ is the distance between two neighboring sites (lattice constant), and the term $\left[1-\delta_{\alpha,\beta}\right]$ ensures that an electron cannot interact with itself. As stated before, the above equation is only valid for the Fock terms, while for the Hartree terms it is the bare interaction (same expression with $\lambda_{TF}\rightarrow\infty$). Then, one can obtain the potential in the HFB approximation using Eq. (\ref{5}).

The electron-electron interaction between the nanowire and the bound charges of the dielectric environment (\emph{extrinsic} interaction) is found in SI \hyperref[Section:SI1]{1}, and it may be implemented following the same procedure by substituting $x\rightarrow \left(i-j\right)/a$. We note that now the term $\left[1-\delta_{\alpha,\beta}\right]$ should not be included since electron $\alpha$ is always inside the nanowire while $\beta$ is outside, in the surrounding medium (or the other way around). Finally, the potential in the HFB approximation can be computed using Eq. (\ref{5}), but now the Fock and the Bogoliubov terms have to be ignored as we argued in the SI \hyperref[Section:SI1]{1}, so that the last four terms of Eq. (\ref{3}) are not considered.

\subsection*{Numerical self-consistent method used to solve the eigenspectrum}
We note that $\hat{H}+e\hat{\phi}^{\rm{HFB}}$ depends on its own eigenvectors (see Eq.(\ref{4})), and thus it has to be solved self-consistently. We solve this problem numerically using the following procedure: in the first iteration of the self-consistent method, we find the density-matrix $\rho$ using the eigenvectors of the Hamiltonian $\hat{H}$ as a test solution. In the next step we obtain a new $\rho$ diagonalizing $\hat{H}+e\hat{\phi}^{\rm{HFB}}$ where the interaction has been obtained using the previous density matrix. In the following steps, the density-matrix is found using a linear combination of the eigenvectors of $\hat{H}+e\hat{\phi}^{\rm{HFB}}$ in the two previous iterations. This is done to introduce damping in the iteration procedure in order to ensure the convergence of the self-consistent method. In each step, we compare the eigen-energies of $\hat{H}+e\hat{\phi}^{\rm{HFB}}$ with those of the previous step. We repeat this procedure until convergence. We consider the iteration has converged  when the difference between both energy-spectra is much smaller than the main energy scale of our problem (i.e. the superconductor gap $\Delta$).

%%%%%%%%%%%%%%%%%%%SI3
\section*{SI 3: Nanowire spectrum including the intrinsic interaction}
\label{Section:SI3}

Here we show that the features studied in the main text (zero-energy pinning and QD formation) remain qualitatively unaltered when including electron-electron interactions $\phi_{int}$  inside the nanowire. The \emph{intrinsic} interaction introduces small quantitative changes in the spectrum, but the qualitative behavior stays the same. Following our previous work \cite{Dominguez:npj17}, we treat this interaction at the mean field level, within the Hartree-Fock-Bogoliubov approximation, and assume a bare Coulomb interaction for the Hartree terms and a screened Coulomb interaction in the quasi-static Thomas-Fermi limit for the Fock terms (see SI \hyperref[Section:SI2]{2} for more details).

\begin{figure}
\includegraphics{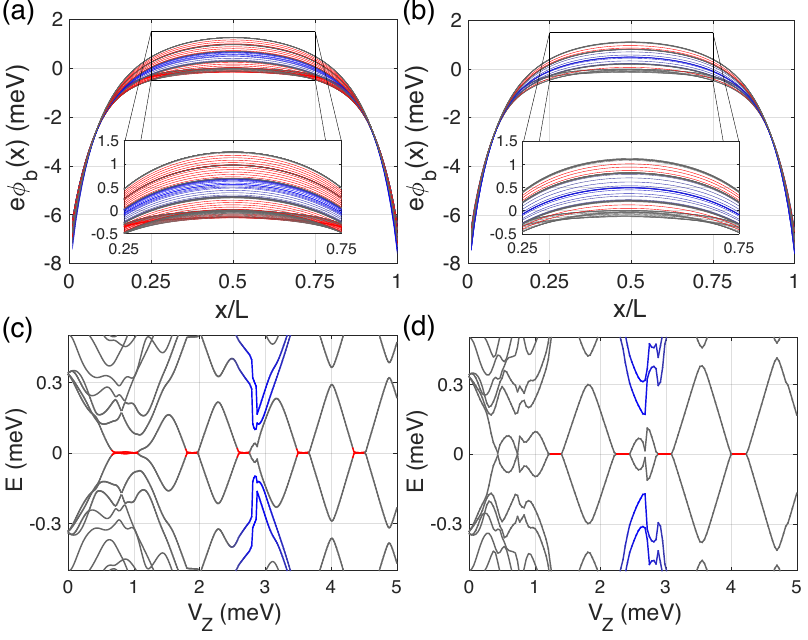}
\caption{Majorana nanowire in the presence of interactions (including
the influence of the bulk normal leads at its ends). Self-consistent induced potential energy
$e\phi_{b}(x)$ along the wire's length for increasing values of the Zeeman splitting ignoring (a) and including (b) the electron-electron interactions inside the nanowire. (c) and (d) are their corresponding energy spectra. Wire parameters are the same as in the main text, and the Thomas-Fermi length is $\lambda_{TF}=L/3$. }
\label{figSI2}
\end{figure}

In Fig. \ref{figSI2} we show the bound charges electrostatic potential along the nanowire and the energy spectrum versus the Zeeman field ignoring (a and c) and including (b and d) the intrinsic interaction. Note that here we do not include the Bogoliubov correction since this term just renormalizes the gap as shown in Ref. \cite{Dominguez:npj17}. In general, both spectra are qualitatively similar and thus we can conclude that the intrinsic interactions do not alter the features studied in the main text. However, there are some quantitative differences. First, the dispersive QD levels approach zero energy at a slightly smaller Zeeman energy as a result of small changes in the induced potential $\phi_b$, as can be seen in Fig. \ref{figSI2}(d). Second, the position of the gap closing and thus, the topological phase transition, shifts to a different magnetic field. This is also a consequence of small changes in $\phi_b$, as well as small changes in the Zeeman energy induced by the Fock terms, that modify the value of the critical Zeeman field through Eq. (4) of the main text. Finally, the energy splitting of the Majoranas is larger due to the renormalization of the Fermi momentum induced by the Fock terms as well.

%%%%%%%%%%%%%%%%%%%SI4
\section*{SI 4: Robustness of the pinning effect}
\label{Section:SI4}

\begin{figure}
\includegraphics{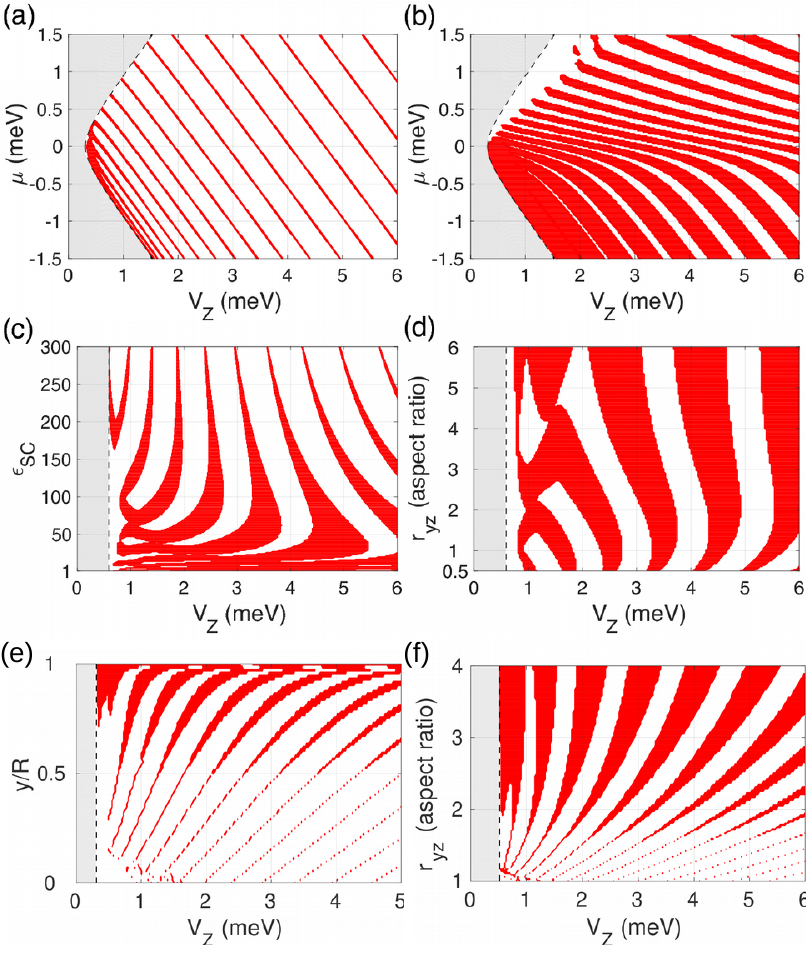}
\caption{Phase diagrams indicating the parameter regions where the Majorana bound states are pinned to zero-energy (in red). In the upper panels the phase diagram is calculated as a function of Zeeman field and chemical potential for the non-interacting case (a) and interacting case(without leads) with $\epsilon_{SC}=100$ (b). The central panels correspond both to the interacting case, but (c) considers variations in the effective dielectric constant of the 
thin superconducting layer, $\epsilon_{SC}$, whereas (d) explores different aspect ratios of the nanowire's section $r_{yz}=W_y/W_z$, where $W_{y,z}$ are the $y$ and $z$ widths of the nanowire faces. In the lower panels we consider perfect screening by the SC shell, $\epsilon_{SC}\rightarrow\infty$, but we vary the distance between the transversal Majorana charge density and the SC shell. In (e) $y/R$ is the position of the Majorana wave function across the nanowire section. When $y=0$ the charge density is at the center of the nanowire, whereas when $y=R$ it is a the opposite face of the SC. In (f) the wave function is fixed at the center of the nanowire section, but we vary its aspect ratio. Here $\mu=0.5$meV and $W_{z}=100$nm, as in the main text.}
\label{figSI3}
\end{figure}

We want to test the validity of our results when varying different parameters of the electrostatic environment. Fig. \ref{figSI3} provides various phase-diagrams indicating the occurrence
of Majorana bound states zero-energy pinning (in red) as a function of the different parameters.
Although we have taken $\mu=0.5$meV for the simulations of the main text, pinning is general for all chemical potentials (within the topological phase), as can be seen in the upper panels of Fig. \ref{figSI3}. As a function of $\mu$ and $V_Z$, the non-interacting lines of (a) corresponding to point-like parity crossings transform into incompressible finite width stripes in the interacting case (b). Pinning regions are bigger for lower chemical potentials and for higher magnetic fields, since the repulsive interaction is larger too. It can also be observed that the onset of the topological phase is different in the interacting system than in the non-interacting one (black dashed line), at least for positive $\mu$. This is because the electrostatic potential renormalizes the chemical potential \cite{Vuik:NJP16} and thus it modifies the value of the bulk critical magnetic field, given in Eq. (\ref{with-leads-critical-field}) of the main text.

However, pinning is not general for all kind of environments. Figure \ref{figSI3}(c) shows the zero energy regions across the $V_Z-\epsilon_{\rm{SC}}$ space (the $\mu-\epsilon_{\rm{SC}}$ diagram exhibits a similar behavior). For $\epsilon_{\rm{SC}}\gtrsim 300$ the pinning plateau width shrinks into points because the electrostatic environment turns into an attractive one. This means that bound charges of the opposite sign arise in the dielectric medium at these large permittivities, so that the entrance of charge at each parity crossing is no longer suppressed. Note that $\epsilon_{\rm{SC}}$ represents the effective SC permittivity of the system composed by a SC thin film (epitaxially grown over the nanowire) of intrinsic permittivity $\epsilon_{\rm{tm}}$, finite width $W_{\rm{SC}}$ and covered by vacuum, as we argue in SI \hyperref[Section:SI1]{1}(IC). For a film width of 8nm, an effective $\epsilon_{\rm{SC}}\gtrsim 300$ corresponds to a SC permittivity of $\epsilon_{\rm{SC}}\gtrsim 15000$, i.e., basically a perfect metal.

The width of the nanowire also plays a role in the appearance of pinned regions. Fig. \ref{figSI3}(d) shows the incompressible regions as a function of $V_Z$ and $r_{yz}$, where $r_{yz}=W_y/W_z$ is the aspect ratio of the nanowire section. When the distance between the SC shell and the opposite side is large (large $r_{yz}$), pinning is bigger. This is because the relative coverage of the wire by the SC shell decreases and so does its attractive contribution.

Finally, if we consider perfect metallic screening by the SC shell, i.e., $\epsilon_{\rm{SC}}\rightarrow\infty$, we can also study the appearance of the pinned regions depending on the distance between the nanowire charge density and the SC shell. In Fig. 4(e) we study the phase diagram as a function of $V_Z$ and the position of the charge density across the nanowire section, $y/R$, where $R$ is the wire's half width. In Fig. 4(f) the position is fixed to the center of the wire, but we vary the wire's aspect ratio. In both cases we observe that, as the Majorana wave functions approaches the SC, the screening effect increases and the pinning disappears \cite{Knapp:17}. However, if the wave function separates from the SC shell, the pinning survives. This may happen when the bottom gate attracts the charge density away from the SC or when the wave function is more spread throughout the wire's section, as for example for higher sub-bands.

\bibliography{beilstein-samuel}
\vspace{3cm}

%%%%%%%%%%%%%%%%%%%%%%%%%%%%%%%%%%%%%%%%%%%%%%%%%%%%%%%%%%%%%%%%%%%%%
%% That's it. Ending the document finishes the article. Happy TeXing!
%%%%%%%%%%%%%%%%%%%%%%%%%%%%%%%%%%%%%%%%%%%%%%%%%%%%%%%%%%%%%%%%%%%%%
\end{document}